\documentclass[footinbib,a4paper,aps,superscriptaddress,reprint,twocolumn,preprintnumbers,amsmath,amssymb,nobalancelastpage,10pt,prx]{revtex4-2}

\usepackage{amsmath}
\usepackage{amssymb}
\usepackage{graphicx}
\usepackage{braket}
\usepackage{yfonts}
\usepackage{dsfont}
\usepackage[colorlinks=true,citecolor=blue,linkcolor=blue,urlcolor=blue]{hyperref}
\usepackage{quantikz}
\usepackage{siunitx}

\begin{document}

\title{Non-Abelian String-Breaking Dynamics on a Qudit Quantum Computer}

\author{Manuel John}
\email{manuel.john@uibk.ac.at}
\author{Keshav Pareek}
\author{Peter Tirler}
\author{Tim Gollerthan}
\author{Michael Meth}
\author{Lukas Gerster}
\affiliation{Universität Innsbruck, Institut für Experimentalphysik, 6020 Innsbruck, Austria}

\author{Peter~Zoller}
\affiliation{Institute for Quantum Optics and Quantum Information, Austrian Academy of Sciences, Innsbruck, 6020, Austria}
\affiliation{Institute for Theoretical Physics, University of Innsbruck, Innsbruck, 6020, Austria}

\author{Daniel González-Cuadra}
\affiliation{Instituto de Física Teórica UAM-CSIC, Universidad Autónoma de Madrid, Cantoblanco, 28049, Madrid, Spain}

\author{Torsten V. Zache}
\affiliation{Institute for Quantum Optics and Quantum Information, Austrian Academy of Sciences, Innsbruck, 6020, Austria}
\affiliation{Institute for Theoretical Physics, University of Innsbruck, Innsbruck, 6020, Austria}

\author{Martin Ringbauer}
\email{martin.ringbauer@uibk.ac.at}
\affiliation{Universität Innsbruck, Institut für Experimentalphysik, 6020 Innsbruck, Austria}

\begin{abstract}

Gauge theories form the foundation of the Standard Model of particle physics. 
These theories can exhibit confinement, where charged particles only occur in bound states, connected by flux strings whose energy grows linearly with separation.
Simulating the real-time dynamics of such strings, including their breaking, remains a major challenge for classical computations and a promising target for quantum simulations. 
While recent quantum simulation experiments explored string-breaking dynamics in abelian lattice gauge theories, non-abelian theories are qualitatively distinct because gauge fields themselves carry charge.
Here, we report the first quantum simulation of genuine non-abelian string-breaking dynamics in a pure SU($2$) lattice gauge theory, where gauge-field self-interactions drive string breaking even in the absence of dynamical matter. 
Our results are obtained on a trapped-ion quantum computer, using native qudit Hilbert spaces to encode truncated gauge fields on a ladder geometry and implement digital Trotter dynamics.
We experimentally study unbreakable and breakable strings generated by fundamental and adjoint static charges, respectively.
We locally resolve string oscillations and coherent string breaking through the creation of gluonic excitations driven by non-abelian plaquette interactions. 
Our work establishes hardware-efficient, problem-tailored qudit simulations as a promising route for accessing non-perturbative dynamics relevant to high-energy physics.

\end{abstract}

\maketitle

\section{Introduction}
Gauge theories form the foundation of the Standard Model of particle physics, where abelian U(1) and non-abelian SU(2), and SU(3) gauge fields mediate the electro-weak and strong interactions. In particular, the latter, described by quantum chromodynamics (QCD) predicts the confinement of SU(3) color charges. As a consequence, quarks---the fundamental carriers of color charge---never appear in isolation, but are bound by gluon fields in colorless composite particles~\cite{gross202350}. As quark--antiquark pairs are separated, the energy stored in the gluon flux tube connecting them increases, eventually leading to the production of additional particles through a process known as string breaking~\cite{Bali_2005}. Although related phenomena arise in lower-dimensional abelian gauge theories such as the Schwinger model~\cite{hebenstreit2013real}, the non-abelian case is qualitatively distinct since gauge fields (gluons) carry charge themselves and self-interact. As a result, in non-abelian SU(N) gauge theories, string breaking can proceed without matter through the formation of gluonic excitations, including glueball-like states~\cite{Pepe_2009}.

Direct observation of these real-time processes is extremely challenging in fundamental particle physics experiments, such as high-energy colliders, and their simulation on high-performance computers from first principles remains difficult~\cite{Troyer_2005}.
Recent quantum simulation experiments~\cite{Altman_2021, Daley_2022} have opened a new pathway to explore the physics of confinement and string breaking with unprecedented spatio-temporal resolution and control~\cite{Bauer_2023, DiMeglio_2023}, by implementing lattice gauge theories (LGTs)~\cite{Montvay_1997} on synthetic quantum devices using, e.g., trapped ions~\cite{Monroe_2021}, neutral atoms~\cite{Gross_2017, Kaufman_2021}, or superconducting qubits~\cite{Kjaergaard_2020}.
So far, however, these experiments have only addressed string breaking in abelian LGTs~\cite{Cochran_2025, Gonzalez-Cuadra2025,  Liu_2025, Alexandrou_2025, crippa2026analysis, De_2024, Cobos_2025, Joshi_2026, Xu_2026}. 

Here, we report the first real-time quantum simulation of non-abelian string breaking dynamics in a pure SU(2) LGT, overcoming the challenge of implementing multi-body plaquette interactions responsible for self-interactions of the non-abelian gauge fields.
Our results are obtained via Trotterized real-time evolution on a digital quantum computer based on trapped-ion qudits~\cite{Ringbauer_2022, Meth2025}, and realize a minimal ladder geometry that reveals the non-abelian nature of these phenomena.
This is achieved by tailoring a recently proposed efficient and scalable truncation scheme for simulating non-abelian LGTs~\cite{Zache_2023_2} to the qudit architecture whose native structure is much better aligned with the physical Hilbert space than a naive qubit embedding.

\begin{figure*}
    \centering{
    \includegraphics[width=\textwidth]{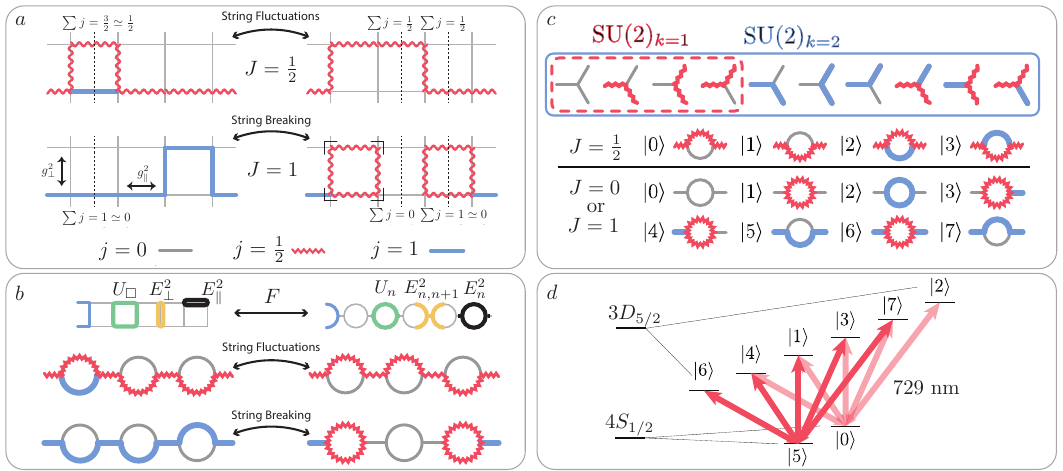}
    \caption{{\bf Encoding SU($2$) gauge fields in trapped-ion qudits.}
    $(a)$ Illustration of non-abelian string fluctuations (top) and string breaking (bottom). A global $\mathbb{Z}_2$ symmetry enforces a superselection rule on the total flux $\sum j \;\mathrm{mod}\; \mathbb{Z}$, indicated by dashed cuts through the chain, fixing it to be either half-integer or integer depending on the boundary charges $J$.  The two sectors are dynamically disconnected, prohibiting string breaking for $J=\frac{1}{2}$, while the integer sector with $J=0,1$ supports gluonic excitations (``gluelumps'' highlighted in the figure). 
    $(b)$ Our quantum simulations are performed in a computational basis that we call a ``bubble chain'', obtained from the standard electric basis via local unitary transformations (so-called ``F-moves''). We indicate the support of the Hamiltonian terms before and after this transformation. 
    $(c)$ We employ an SU($2$)$_k$ approximation of the SU($2$) LGT, truncating electric fluxes to $j=0, \frac{1}{2}, 1, \dots, \frac{k}{2}$, where Gauss' law manifests through fusion rules as a constraint at vertices where three flux lines meet. We focus on the smallest truly non-abelian case $k=2$; in contrast, $k=1$ (dashed box) yields fusion rules equivalent to an abelian $\mathbb{Z}_2$ LGT. In the bubble-chain basis, the two flux sectors introduced in $(a)$ map to local Hilbert spaces of dimension four (half-integer sector) and eight (integer sector), enabling an efficient encoding with one qudit per plaquette. 
    $(d)$ In our experiments, single qudits are implemented using internal states of individual $^{40}$Ca$^+$ ions, encoded in the $4S_{1/2}$ and $3D_{5/2}$ manifolds and coherently coupled via a narrow-linewidth 729\,nm laser.}
    \label{fig:overview}}
\end{figure*}

In particular, we realize the dynamics of both unbreakable strings connecting fundamental charges and breakable strings connecting ``adjoint'' charges. We monitor their coherent quantum evolution and observe simple oscillations of the unbreakable string, contrasted with non-abelian string-breaking dynamics in the adjoint sector, including the production of \emph{gluonic} excitations driven by plaquette interactions.

\begin{figure*}[ht!]
    \centering
    \includegraphics[width=\textwidth]{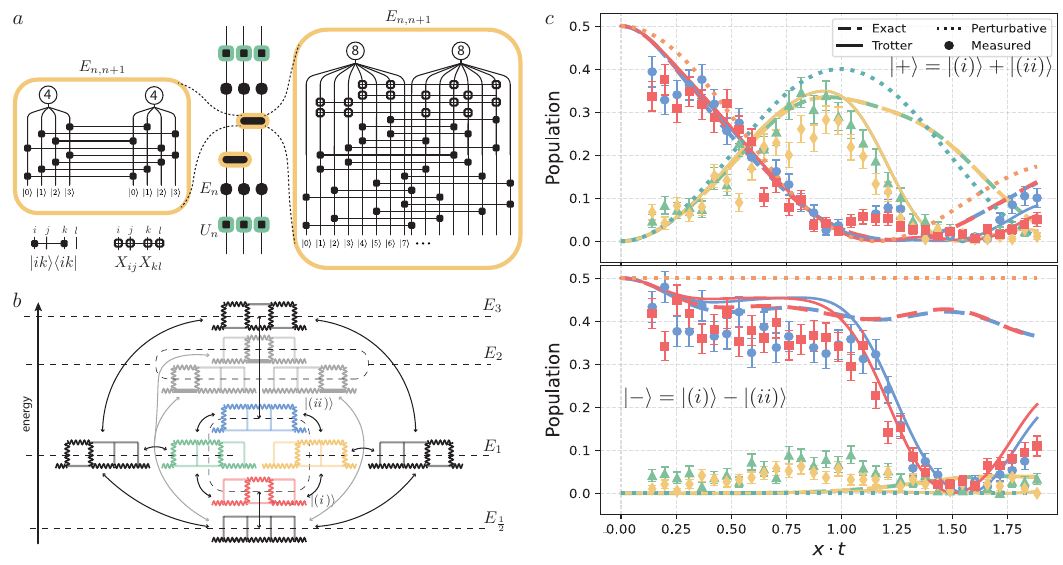}
    \caption{\textbf{Non-abelian string fluctuations.} We consider strings with boundary charges $J=1/2$, which constrains the dynamics to an unbreakable string sector.
    (a) We simulate the real-time dynamics of the system through a second-order Suzuki-Trotter time evolution. In the figure, we show the circuit corresponding to one Trotter step, where  we depict the entangling layer explicitly as acting between sublevels of neighboring qudits --- on the left for the case of string fluctuations leading to time evolutions shown in (c) and on the right for the case of string-breaking simulations presented in Fig.~\ref{fig:string_breaking_1}. (b) For $x=0$, the first excited manifold consists of $N_\square(N_\square+1)/2$ degenerate states corresponding to local perpendicular displacements of the string. In the regime $x \ll g^2_{\perp, \parallel}$, the dynamics are restricted to this manifold and are well described by an effective hopping model (see Eq.~\ref{eq:fermionic_hopping}) with amplitude $J_\mathrm{eff} = -x/\sqrt{2}$. We solve this effective model analytically, with the resulting populations indicated as \emph{perturbative} in (c). The level scheme also indicates nearby string configurations, with arrows highlighting plaquette-induced transitions connecting them.
    (c) We experimentally prepare superpositions of two such states: (i) a perpendicular displacement at the center and (ii) a perpendicular displacement extending across the system which we evolve in time for the parameters $x=0.3$, $g^2_{\perp}=1$, $g^2_{\parallel}=1.5$. The symmetric $\ket{+}$ superposition shows constructive interference, resulting in coherent fluctuations of the string between neighboring configurations. In contrast, the antisymmetric $\ket{-}$ superposition experiences destructive interference and suppresses these fluctuations, leading to a near-freezing of the initial state. The measured populations directly reveal interference-controlled coherent string fluctuations within the constrained Hilbert space.
    }
    \label{fig:string_fluctuations}
\end{figure*}

\section{Model and qudit implementation}
Our target is a pure SU($2$) LGT described by the Hamiltonian
\begin{align}{\label{eq:H}}
H = \frac{g_\perp^2}{2} \sum_{\ell \in \perp} E_\ell^2 + \frac{g_\parallel^2}{2} \sum_{\ell \in \parallel} E_\ell^2 - x \sum_\square \mathcal{U}_\square \;,
\end{align}
where gauge fields live on the links $\ell$ of a two-dimensional square lattice, which we consider in the extreme anisotropic limit of a quasi-1D chain of plaquettes, see Fig.~\ref{fig:overview}(a).
Despite this simplification, the model hosts a variety of phenomena characteristic of non-abelian gauge theories, driven by the interplay of electric energy $E_\ell^2$, which we control independently in longitudinal ($g^2_\parallel$) and perpendicular ($g^2_\perp$) directions, and magnetic plaquette interactions $\mathcal{U}_\square$ (controlled by $x$).

We employ a $q$-deformed regularization of SU(2)$_k$ gauge fields, with $k=2$ being the smallest choice that features true non-abelian behavior~\cite{Zache_2023_2, Hayata_2023_1}.
This regularization truncates the infinite-dimensional local Hilbert space associated with SU($2$) gauge fields to a finite dimension of $k+1$, while preserving the algebraic structure of the plaquette operators in Eq.~\eqref{eq:H}, which enables efficient and scalable quantum algorithms in higher dimensions~\cite{Zache_2023_2} and more general gauge groups such as SU($3$)~\cite{hayata2023q}.
As illustrated in Fig.~\ref{fig:overview}(c), we developed an optimized hardware-aware encoding that represents the gauge field configurations with a single qudit per plaquette, each realized by one trapped ion, with local dimension at most \(d{=}8\), see Fig.~\ref{fig:overview}(d). This is achieved by first transforming the plaquette ladder into a bubble chain via a basis change implemented through a sequence of so-called $F$-moves, see Fig.~\ref{fig:overview}(b) and Refs.~\cite{Gils_2009,hayata2025floquet}. In our encoding, this transformation turns plaquette interactions into single-qudit operators, while the electric term operators can be realized with at most two-qudit entangling gates---significantly reducing the overall gate cost compared to qubit-based encodings~\cite{Gonzalez-Cuadra_2022, Zache_2023_1, Popov_2023, Silvi2024, Kurkcuoglu_2024, Ballini_2025, Meth2025, Gaz_2025, Jiang_2025, Joshi_2025_1, Joshi_2025_2, Popov_2025} (see Appendix~\ref{sec:qubit-vs-qudit-embedding}).

In the following, we demonstrate our approach in a minimal setup with $N_\square = 3$ plaquettes on our universal trapped-ion qudit quantum computer. While previous experiments have explored real-time dynamics of two-dimensional SU($2$) LGTs using related or alternative regularization schemes~\cite{Klco_2020, Rahman_2022, hayata2026, Chen_2026}, these works either relied on stronger truncations that effectively reduce the model to an abelian one or were limited to smaller system sizes.
Here, we overcome these limitations, which is essential for observing genuine non-abelian string-breaking dynamics.

Throughout the rest of this article, we study the real-time evolution of this SU(2)$_2$ LGT by implementing digital Trotter circuits that capture the dynamics generated by the Hamiltonian in Eq.~\eqref{eq:H} (see Appendix~\ref{sec:hamiltonian}). 
Crucially, our approach preserves a global $\mathbb{Z}_2$ symmetry, associated with the center of SU(2)~\cite{Pepe_2009}, which separates the dynamics into distinct flux-parity sectors. 
We access these sectors through the static boundary charges shown in Fig.~\ref{fig:overview}(a): fundamental charges, which act as sources of $J=1/2$ flux and generate half-integer-flux strings, and adjoint charges, which act as sources of $J=1$ flux and generate strings in the integer-flux sector.

\section{String fluctuations}
We first consider boundary conditions corresponding to static charges that inject fundamental flux, \(J=1/2\), as shown in Fig.~\ref{fig:overview}(a). 
The electric energy of a single link carrying SU(2) representation \(j\) is \(j(j+1)g_{\alpha}^2/2\), where \(\alpha=\parallel,\perp\) denotes whether the link is parallel or perpendicular to the string direction. For \(x=0\), the ground state is a straight, unbreakable string configuration \(\ket{S_{1/2}}\) connecting the two boundaries, with energy \(E_{1/2}=\sigma_{1/2}N_\square\), where \(\sigma_{1/2}=\tfrac{3}{8}g_\parallel^2\) is the bare string tension. 
Turning on the plaquette term ($x>0$) does not break this string, but induces local transverse fluctuations, generating excited string configurations, with excess energy \(\sim 2\times \tfrac{3}{4}(g_\perp^2/2)\) when the string is displaced away from the straight path on exactly two perpendicular segments, see Fig.~\ref{fig:string_fluctuations}(b) with energy manifold $E_1$.

This picture becomes particularly transparent in the perturbative regime \(x \ll g_\perp^2, g_\parallel^2\), where the plaquette term weakly couples these excited string configurations while their energies are set predominantly by the electric terms. In this limit, the low-energy dynamics is captured by the effective Hamiltonian
\begin{align}
    H_\text{eff}
    =
    J_\text{eff}
    \sum_{\langle \ell_\perp,\ell'_\perp\rangle}
    \left(
    s_{\ell_\perp}^\dagger s_{\ell'_\perp}
    + \text{h.c.}
    \right),
\label{eq:fermionic_hopping}
\end{align}
where \(s_{\ell_\perp}^\dagger\) and \(s_{\ell_\perp}\) create and annihilate a local transverse displacement of the string, and \(J_\text{eff}=-x/\sqrt{2}\) is the effective hopping amplitude induced by the plaquette term. Eq.~\eqref{eq:fermionic_hopping} thus describes the coherent propagation of transverse string excitations along the ladder. The dynamics within this restricted manifold can be solved analytically (see Appendix~\ref{sec:analytical_solution}), providing a minimal description of the oscillations between the dominant fluctuating string configurations.

To test this effective picture of coherent string fluctuations in the non-breaking $J=1/2$ sector, we experimentally prepare coherent superpositions of two string configurations with different relative phases, see Fig.~\ref{fig:string_fluctuations}(b). The dynamics are implemented digitally using two second-order Trotter--Suzuki steps, with the effective evolution time set by the step size; one Trotter step is shown in Fig.~\ref{fig:string_fluctuations}(a), and the corresponding Hamiltonian in the computational basis and circuit decomposition are detailed in Appendix~\ref{sec:hamiltonian}. To interpret the data, we compare the measurements both to exact numerical time evolution under the full Hamiltonian and to the analytical solution of the perturbative effective model introduced in Eq.~\eqref{eq:fermionic_hopping}.

In Fig.~\ref{fig:string_fluctuations}(c), we observe a strong dependence on the relative phase: symmetric superpositions exhibit constructive interference and coherent fluctuations within the first excited manifold, whereas antisymmetric superpositions show strongly suppressed dynamics due to destructive interference between competing transition pathways. This suppression is therefore not merely symmetry-protected, but arises from interference in the interacting system.

The perturbative effective model provides a minimal description of this mechanism and qualitatively describes the phase-sensitive nature of the dynamics. Its restricted Hilbert space, however, contains only the resonantly coupled configurations in the first excited manifold and therefore omits leakage into higher-energy configurations generated by the full Hamiltonian. Consequently, small deviations from the exact full-Hamiltonian evolution become visible as $x t$ approaches unity. The Trotterized dynamics agree more closely with the exact evolution on this timescale, demonstrating that the implemented circuit captures the relevant full-Hamiltonian dynamics beyond the reduced perturbative description. At later times, exact and Trotterized results separate as Trotter errors accumulate, while the experimental data continue to follow the Trotterized simulation closely. This confirms that the measured dynamics reflect the intended digital quantum simulation, including effects beyond the minimal effective model.

\begin{figure*}
    \centering
    \includegraphics[width=\textwidth]{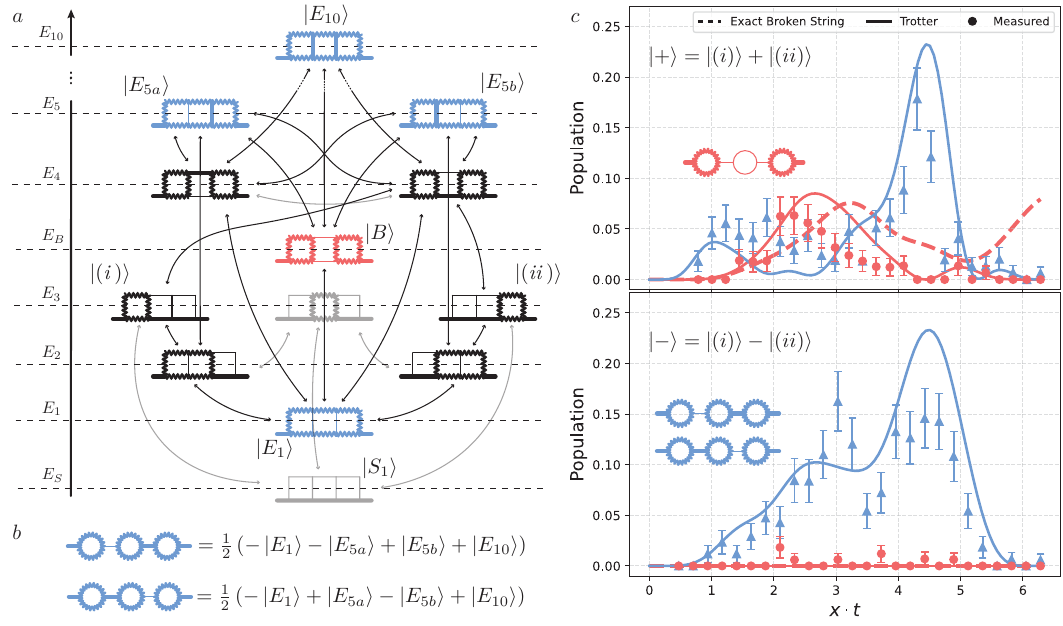}
    \caption{\textbf{Non-abelian string breaking}. We experimentally observe dynamical string breaking of a $J=1$ flux string via coherent population transfer into a broken-string configuration. (a) In contrast to the unbreakable $J=1/2$ case, two competing configurations exist: an unbroken string $\ket{S_1}$ with finite string tension, and a broken string $|B\rangle$ in which the flux is screened by the creation of two glueballs. A transition between $\ket{S_1}$ and $\ket{B}$ is enabled by the non-abelian fusion rule $\frac{1}{2} \otimes \frac{1}{2} = 0 \oplus 1$.
    The level scheme shows the most relevant string configurations, with arrows indicating plaquette-induced transitions (the energies are scaled arbitrarily, but their order correctly reflects the experimental values of $g^2_{\perp, \parallel}$).
    (b) While $\ket{S_1}$ and $\ket{B}$  map one-to-one onto the computational bubble basis (defined in Fig.~\ref{fig:overview}), intermediate states instead appear as superpositions in the encoded bubble basis.
    These are related to the four intermediate configurations that necessarily precede string breaking, although there are several competing paths of at least third order which make string breaking a highly non-linear process.
    (c) To probe coherence, we prepare superpositions of two unbroken string states dressed by a glueball, initialized (i) on the left or (ii) on the right side of the ladder, and evolve the system in time with $x=1.0$, $g^2_{\perp}=0.8$, $g^2_{\parallel}=2.0$. We show the combined summed population of the two dominant intermediate configurations (in the computational basis) and the population of the broken-string state.
    In the $\ket{+}$ case (top), the broken string reaches populations of up to $\sim 8\%$, while it remains suppressed in the $\ket{-}$ case (bottom) due to destructive interference.
    These results constitute a first observation of dynamical string breaking of a non-abelian SU($2$) flux string in a coherent quantum simulation based on up to eight-dimensional qudits.}
    \label{fig:string_breaking_1}
\end{figure*}

\section{String-breaking dynamics}
We now turn to strings with higher adjoint boundary charges, injecting flux $J=1$, see Fig.~\ref{fig:overview}(a). In this case, the non-abelian nature of SU($2$) permits flux strings to break through the production of glueball pairs, ultimately screening the static charges in configurations often called \emph{gluelumps}. This is a fundamentally non-abelian dynamical process that is absent in abelian LGTs.

Consider first the system without magnetic interactions ($x=0$), where two ground state candidates emerge, as shown in Fig.~\ref{fig:string_breaking_1}(a): an unbroken string $|S_1\rangle$ with energy $E_S = \sigma_1 N_\square$ and string tension $\sigma_1(g_\parallel) = g_\parallel^2$; and a broken string $|B\rangle$, fully screened by two glueballs, with energy $E_B = \frac{3}{2}(g_\perp^2 + g_\parallel^2)$.
As a result, $E_S = E_B$ predicts a resonance for string breaking at
\begin{align}
\left(\frac{g_\perp^2}{g_\parallel^2}\right)_\text{SB} = \frac{2}{3}N_\square -1 \;,
\label{eq:resonance_condition}
\end{align}
which occurs for sufficiently large strings ($N_\square \ge 3$).
Intuitively, string-breaking occurs when the string tension $\sigma_1(g_\parallel)$ increases beyond the bare glueball mass $M_B (g_\perp,g_{\parallel}) = E_B/2$. In the experiment, we can control these contributions independently by varying $g_\perp^2$ and $g_\parallel^2$. We emphasize that the fact that strings of integer flux $j$ can break by the production of glueballs is a direct consequence of the non-abelian nature of SU($2$), here specifically the fusion rule $\frac{1}{2} \otimes \frac{1}{2} = 0 \oplus 1$, which is absent at the strongest truncation $k = 1$ and first emerges at $k=2$.

Similarly to the $J=1/2$ case, we study the coherent dynamics of $J=1$ strings by preparing symmetric and antisymmetric superpositions (with respect to spatial reflection about the ladder center) of glueball-dressed strings localized on the left and right of the ladder, see Fig.~\ref{fig:string_breaking_1}(c). The symmetric $\ket{+}$ superposition leads to constructive interference, enabling population transfer into the broken-string configuration, whereas the antisymmetric $\ket{-}$ suppresses this process via destructive interference. In contrast to the $J=1/2$ case, this spatial symmetry is respected by the full Hilbert space. Time evolution is again implemented digitally using two second-order Trotter--Suzuki steps, with the effective evolution time set by the Trotter step size. Despite the highly non-linear nature of string breaking---arising from multi-step couplings between configurations evident in the level structure in Fig.~\ref{fig:string_breaking_1}(a)---the digital dynamics capture the broken-string population in agreement with exact simulations. Overall, the evolution highlights how interference controls the onset of string breaking and demonstrates coherent, phase-sensitive non-abelian dynamics in trapped-ion qudit simulations with local dimension up to \(d{=}8\).

\begin{figure}[t]
    \centering
    \includegraphics[width=\columnwidth]{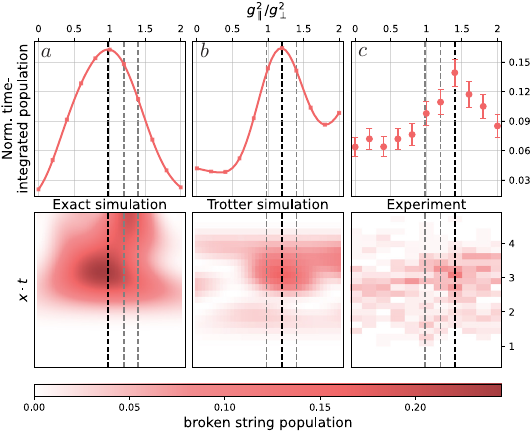}
    \caption{\textbf{Tunable resonance of non-abelian string breaking.}
    Varying the ratio of coupling strengths $g_\parallel^2/g_\perp^2$ tunes the relative energies of the unbroken and broken string configurations, $\ket{S_1}$ and $\ket{B}$, and hence the resonance condition for string breaking. We study the real-time dynamics of a string initialized in $\ket{S_1}$, showing the population of the broken-string configuration $\ket{B}$ as a function of $g_\parallel^2/g_\perp^2$ and evolution time. The bottom panels display simulated and measured populations, while the top panels show the corresponding time-integrated population, making the resonance profile more apparent.
    In the limit $x \ll g^2_{\parallel,\perp}$, resonance occurs at energetic degeneracy, which for $N_\square=3$ yields $g_\parallel^2/g_\perp^2 = 1$. For the experimental parameters, including a stronger magnetic term $(x=0.78, g_{\parallel}^2=2.0)$, the resonance is broadened. The trotterized dynamics in column (b) exhibit a systematic shift of the peak position relative to the exact result (a) due to Trotter errors, while the experimental data in column (c) shows a further shift that we attribute to experimental imperfections. 
    Despite these shifts, all three approaches consistently display enhanced population transfer over a finite region of $g_\parallel^2/g_\perp^2$ around unity, signaling resonant string breaking. The peak positions of the exact, trotterized, and experimental results are indicated by vertical dashed lines. The time-integrated curves are normalized over the experimentally sampled $g_\parallel^2/g_\perp^2$ values, such that they reflect the relative weight of the resonance feature, which is consistent across exact, trotterized, and experimental dynamics.}
    \label{fig:string_breaking_2}
\end{figure}

Building on the interference-controlled dynamics of Fig.~\ref{fig:string_breaking_1}(c), we now investigate the parameter dependence of string breaking by varying the parallel coupling strength $g_\parallel^2$. Figure~\ref{fig:string_breaking_2} shows the population dynamics of $\ket{B}$ as a function of $g_\parallel^2/g_\perp^2$ and the evolution time $xt$, for a string initialized in the unbroken configuration $\ket{S_1}$. In addition to the time-resolved populations, we consider the corresponding time-integrated population of the broken-string state $\ket{B}$, which provides a direct measure of the resonance profile, normalized across the explored coupling values for comparison.

In the perturbative regime, \(x \ll g_{\parallel,\perp}^2\), string breaking is resonantly enhanced when the unbroken and broken string configurations become energetically degenerate. For \(N_\square = 3\), this occurs at \(g_\parallel^2/g_\perp^2 = 1\), as per Eq.~\eqref{eq:resonance_condition}. Although the experimental parameters \((x=0.78,\, g_\parallel^2=2.0)\) lie outside this limit, the exact, Trotterized, and experimental dynamics each show enhanced population transfer over a finite range of $g_\parallel^2/g_\perp^2$ around unity, signaling resonant string breaking. In our regime, the resonance is broadened relative to the perturbative expectation. Moreover, relative to the exact dynamics, the Trotterized simulation shows a systematic shift of the resonance peak due to Trotter errors, while the experimental data exhibits a further shift that we attribute to experimental imperfections. The time-integrated populations make this resonance profile particularly apparent and show overall consistency between exact, Trotterized, and experimental dynamics.

\section{Discussion and Outlook}
We have reported the first real-time quantum simulation of non-abelian string-breaking dynamics in a pure SU(2) lattice gauge theory. Using a trapped-ion qudit quantum computer with local Hilbert spaces of dimension up to \(d{=}8\), together with an efficient SU(2)\(_2\) truncation and a hardware-optimized gauge-field encoding, we realized both unbreakable half-integer-flux strings and breakable integer-flux strings on a minimal plaquette ladder. The measured dynamics reveal coherent, phase-sensitive string fluctuations and purely gluonic screening of adjoint charges, demonstrating the key role of non-abelian plaquette interactions in enabling string breaking without dynamical matter.

The present experiment should be viewed as a proof-of-principle demonstration of real-time non-abelian string-breaking dynamics.
Aiming at a regime where new gauge-theory phenomena are expected to emerge, we face important challenges, including small system sizes, finite Trotter step errors, experimental gate imperfections, and the restricted gauge-field truncation. 
Overcoming these challenges will require improved qudit control, more efficient real-time evolution protocols, and extensions to larger local Hilbert spaces and longer plaquette ladders.

More broadly, our results highlight qudit-based quantum processors as a natural platform for simulating non-abelian LGTs, where the physical degrees of freedom are intrinsically higher-dimensional. The combination of gauge-tailored truncations and hardware-efficient qudit encodings provides a concrete route toward larger SU(2)\(_k\) simulations, higher-dimensional geometries, and eventually more complex gauge groups such as SU(3). These developments will open the door to controlled studies of non-perturbative real-time dynamics in regimes beyond the reach of exact classical computation.

\section*{Acknowledgements}
D.G.-C. acknowledges financial support through the Ramón y Cajal Program (RYC2023-044201-I), financed by MICIU/AEI/10.13039/501100011033 and by the FSE+.
This research was funded by the European Research Council (QUDITS, 101039522), and by the European Union under the Horizon Europe Programme—Grant Agreement 101080086—NeQST. T.V.Z.~is supported by an ERC Starting grant (QS-Gauge, 101220401). 
Funded by the European Union. Views and opinions expressed are however those of the author(s) only and do not necessarily reflect those of the European Union or the European Research Council Executive Agency. Neither the European Union nor the granting authority can be held responsible for them.
We also acknowledge support by the IQI GmbH. 

\paragraph*{Note added.--} While completing this work, we were made aware of Ref.~\cite{Xu_2026} by K.~Xu et al., studying the dynamics of glueballs in an abelian Z2 lattice gauge theory on a trapped-ion quantum computer.

\bibliography{bibliography}

\newpage
\section*{Appendices}
\appendix

\section{Experimental setup}

Experiments were performed on linear chains of three $^{40}$Ca$^{+}$ ions confined in a macroscopic linear Paul trap. The ions are prepared close to the motional ground state using Doppler cooling, followed by resolved sideband cooling \cite{Schindler_2013}, resulting in a mean phonon occupation of $\bar{n}_y \approx 0.03(1)$ of the highest radial center-of-mass mode with a secular trap frequency of $\omega_y = 2\pi \cdot \SI{2.870}{\mega\hertz}$.

Quantum information is encoded in Zeeman sublevels of the $4S_{1/2}$ ground state and the metastable $3D_{5/2}$ manifold. An external magnetic field of 6.38 G lifts the degeneracy of these levels, yielding two $4S_{1/2}$ and six $3D_{5/2}$ Zeeman sublevels that can be spectroscopically addressed, providing up to eight distinct internal states per ion. Coherent operations between the $S_{1/2}$ and $D_{5/2}$ manifolds are driven using a narrow-linewidth \SI{729}{\nano\meter} laser stabilized to a high-finesse optical cavity. The coherence time between the $4S_{1/2}(m_j = -1/2)$ level and any of the other levels ranges from 50 to several hundred milliseconds. The natural lifetime of the $D_{5/2}$ orbitals is approximately \SI{1.17}{\second} \cite{Barton_2000}.

State detection is performed via state-dependent fluorescence on the $4S_{1/2} \leftrightarrow 4P_{1/2}$ optical dipole transition \cite{Schindler_2013}. Due to the linewidth of this transition, the two $S_{1/2}$ ground states are indistinguishable \cite{Ringbauer_2022}. However, this does not limit coherent control within the eight-dimensional qudit manifold. 

Individual ions are addressed by tightly focusing the \SI{729}{\nano\meter} beam through a high-NA objective aligned perpendicular to the ion string, resulting in residual addressing crosstalk between neighboring ions below \SI{1}{\percent} in optical intensity. Randomized benchmarking of resonant two-level Clifford operations with average duration of \SI{50}{\micro\second} yields a fidelity of \SI{99.96(15)}{\percent}. Entangling interactions between ions are implemented using the Mølmer–Sørensen (MS) interaction \cite{MolmerSorensen_1999}, applied with a detuning of approximately $2\pi \cdot \SI{3}{\kilo\hertz}$ from the highest y-radial motional mode, resulting in maximally entangling gate durations of $t = \SIrange{310}{315}{\micro\second}$ and gate fidelities of \SI{99.2(5)}{\percent}. To minimize calibration overhead, we drive entangling operations on a single transition ($\ket{4S_{1/2}, m_j=-1/2} \Leftrightarrow \ket{3D_{5/2}, m_j=-1/2}$) at the cost of at most three additional local operations per ion that bring the correct states into the interacting manifold. Correcting for local phases, the native entangling Mølmer–Sørensen interaction, where $\ket{i},\ket{j}$ denote the two interacting states of one ion and $\ket{k},\ket{l}$ those of the other ion, is generated by the following operator
\begin{equation}
    \text{X}_{ij}\text{X}_{kl} = \ket{ik}\bra{jl} + \ket{il}\bra{jk} + \ket{jk}\bra{il} + \ket{jl}\bra{ik}
\label{eq:xx_interaction}
\end{equation}
or, through local rotations, equivalently by
\begin{equation}
    \text{Z}_{ij}\text{Z}_{kl} = \ket{ik}\bra{ik} - \ket{il}\bra{il} - \ket{jk}\bra{jk} + \ket{jl}\bra{jl}
\label{eq:zz_interaction}
\end{equation}
The light shifts induced on the 6 spectator levels due to the MS pulses range from \SIrange{0.4}{2}{\kilo\hertz} with respect to the $4S_{1/2}(m_j = -1/2)$ level, corresponding to a phase of \SIrange{30}{140}{\degree} accumulated per fully-entangling gate. \\

\section{Hamiltonian in the computational basis}
\label{sec:hamiltonian}
The plaquette ladder can be converted into a bubble chain [Fig.~\ref{fig:overview}(b)] by applying a sequence of local basis transformations called $F$-moves~\cite{Gils_2009,Zache_2023_2,hayata2025floquet}. These are unitary operators that transform between different ``point-splitting'' bases for the SU($2$) LGT on a square lattice. In the bubble basis, the electric operator decomposes into local terms $E_n$ acting on bubble $n$ and nearest-neighbor interactions $E_{n,n+1}$ between adjacent bubbles. Also the plaquette operator $U_{\square}$, which acts on four links in the original basis, becomes a local operator $U_n$ acting on a single bubble.

For boundary charge $J = 1/2$, in the qudit basis defined in Fig.~\ref{fig:overview}(c), the plaquette interaction is given by
\begin{align}
     U_n = \frac{1}{\sqrt{2}}
    \begin{pmatrix}
	0 & -1 & 0 & 1 \\
	-1 & 0 & 1 & 0 \\
	0 & 1 & 0 & -1 \\
	1 & 0 & -1 & 0 
    \end{pmatrix}.
\end{align}
The electric term is local in the bulk, where each bubble carries an energy
\begin{align}
    E^2_{ n} 
    = \frac{g^2_{\parallel}}{2} \, \text{diag} 
    \begin{pmatrix}
    \tfrac{3}{4} & \tfrac{3}{4} & \tfrac{11}{4} &  \tfrac{11}{4}
    \end{pmatrix} \;,
\end{align}
while at the boundaries, flux continuity across the edges of the plaquette ladder leads to an additional energy contribution. As a result, the outer bubbles are described by
\begin{align}
E^2_{0} = E^2_{N-1}
&=
E^2_{n}
+
\frac{g^2_{\perp}}{2} \, \text{diag} 
    \begin{pmatrix}
    \tfrac{3}{4} & 0 & \tfrac{3}{4} &  2
    \end{pmatrix} \;.
\end{align}
The nearest-neighbor electric interaction $E^2_{n,n+1}$ acts on two adjacent bubbles and reads
\begin{align}
E^2_{n,n+1}
= \frac{g^2_{\perp}}{2}\Bigg[
&\frac{3}{4}\Big(
|01\rangle \langle 01| + |10\rangle \langle 10|
+ |03\rangle \langle 03| + |30\rangle \langle 30| \nonumber\\
&
+ |12\rangle \langle 12| + |21\rangle \langle 21|
+ |23\rangle \langle 23| + |32\rangle \langle 32|
\Big) \nonumber\\
+2&\Big(
|02\rangle \langle 02| + |20\rangle \langle 20|
+ |13\rangle \langle 13| + |31\rangle \langle 31|
\Big)
\Bigg].
\end{align}

For boundary charge $J=1$, we work with $N$ eight-dimensional qudits, as defined in Fig.~\ref{fig:overview}(c).
The plaquette term is a local operation on each ion,
\begin{equation}
    U_{n} = \big[ |1\rangle (\langle 0| + \langle 2|) - |6\rangle (\langle 5| + \langle 7|) \big] + \mathrm{h.c.},
\end{equation}
while the local electric term on the inner bubbles carries energy
\begin{align}
E^2_{n} = \frac{g^2_{\parallel}}{2} \, \text{diag} 
    \begin{pmatrix}
    0 &
    \tfrac{3}{2} &
    4&
    \tfrac{3}{2} & 
    \tfrac{3}{2} & 
    2 &
    \tfrac{3}{2} &
    2
    \end{pmatrix} \;,
\end{align}
and outer bubbles again carry higher energy due to flux continuity at the edges of the plaquette ladder,
\begin{align}
E^2_{0} = E^2_{N-1}
&=
E^2_{n}
+
\frac{g^2_{\perp}}{2}  \, \text{diag} 
    \begin{pmatrix}
    0 &
    \tfrac{3}{4} &
    2&
    \tfrac{3}{4} & 
    \tfrac{3}{4} & 0 &  \tfrac{3}{4}
    & 2
    \end{pmatrix} \;.
\end{align}
The electric interaction further contains a partially non-local contribution acting on neighboring bubbles,
\begin{align}
E_{n,n+1}^2 &= \frac{g^2_{\perp}}{2} \times\\
\Bigg[
&\frac{3}{4}\big(\ket{01} \bra{01} + \ket{03} \bra{03} + \ket{21}\bra{21} + \ket{23}\bra{23} \nonumber \\
&+ \ket{10}\bra{10} + \ket{12}\bra{12} + \ket{40}\bra{40} + \ket{42}\bra{42} \nonumber \\
&+  \ket{65}\bra{65} + \ket{67}\bra{67} + 
\ket{35}\bra{35} + \ket{37}\bra{37}\nonumber \\
&+ \ket{54}\bra{54} + \ket{56}\bra{56} + \ket{74}\bra{74} + \ket{76}\bra{76} \big)  \nonumber \\[10pt]
+ &2\big(\ket{02}\bra{02} + \ket{20}\bra{20} + \ket{57}\bra{57} + \ket{75}\bra{75} \big) \nonumber \\[10pt]
+&\ket{11}\bra{11} + \ket{13}\bra{13} + \ket{41}\bra{41} + \ket{43}\bra{43} \nonumber \\
+&\ket{34}\bra{34} +\ket{36}\bra{36} + \ket{64}\bra{64} + \ket{66}\bra{66} \nonumber \\[10pt]
+&\big(
\ket{11}\bra{34} + \ket{13}\bra{36} + \ket{41}\bra{64} + \ket{43}\bra{66}
 \nonumber\\
&\quad + \mathrm{h.c.}\big)
\Bigg]
\label{eq:nonlocal_e_breaking}
\end{align}

\section{Trotterized time evolution}
Using the above tools we implement $m$ steps of a second-order Suzuki-Trotter approximation with the theoretical 
error scaling $O(t^3/m^2)$, where each step is defined by 
\begin{align}
U_{U}\!\left(\tfrac{t}{2}\right)
U_{E_n}\!\left(\tfrac{t}{2}\right)
&U_{E_{n,n+1}}(t)
U_{E_n}\!\left(\tfrac{t}{2}\right)
U_{U}\!\left(\tfrac{t}{2}\right),\\
U_{U}(t) &= \prod_{n=0}^{N-1} e^{-ixU_n t},\\
U_{E_\mathrm{n}}(t) &= \prod_{n=0}^{N-1} e^{-iE_n^2 t},\\
U_{E_\mathrm{n,n+1}}(t) &= \prod_{n=0}^{N-2} e^{-iE_{n,n+1}^2 t}.
\end{align}
This results in the circuit structure shown in Fig.~\ref{fig:string_fluctuations}(a), where longer evolution times correspond to larger gate angles (longer durations). For Mølmer–Sørensen entangling gates, the interaction angle can be increased by scaling the laser power; beyond a certain threshold, multiple gates must be concatenated to reach larger angles.

\section{Physicality}
Gauss' law restricts the dynamics to a physical subspace, determined by the SU(2)$_k$ fusion rules, which can be exploited to reduce the number of entangling gates. In general, an interaction of the form Eq. \eqref{eq:xx_interaction} generates four transition terms, only a subset of which act within the physical subspace.

To illustrate this, consider a two-qubit system where the states $\ket{01}$ and $\ket{10}$ are unphysical. Then
\begin{equation}
    \text{X}_{01}\text{X}_{01} = \ket{00}\bra{11} + \ket{01}\bra{10} + \ket{10}\bra{01} + \ket{11}\bra{00}
\label{eq:xx_0101}
\end{equation}
but only the first and last terms act nontrivially. The remaining contributions can therefore be neglected when acting on states obeying Gauss' law.

The same mechanism applies in our qudit system. For example, while a decomposition of a term such as $\ket{43}\bra{66} + h.c.$ would formally require combining $X_{46}X_{36}$ and $Y_{46}Y_{36}$ interactions to cancel unwanted contributions, these corresponding ``odd'' terms (e.g. $\ket{46}\bra{63}$ + h.c.) lie outside the physical subspace. As a result, these contributions effectively vanish, and the desired interaction can be implemented using a single MS gate. An equivalent argument holds for the $Z_{ij}Z_{kl}$-type interactions.

Furthermore, some errors that take the system outside the physical subspace can be filtered via post-selection. These errors are observed to increase approximately linearly with evolution time/MS gate count (see 
Fig.~\ref{fig:physicality}) and reach a maximum of \SI{27}{\percent} in our longest time evolutions. In contrast, for the string fluctuation simulations all state combinations are physical, and this mechanism is not available.

We also take advantage of physicality in combination with a global phase in the entangling layer --- the explicit decomposition of $E_{n,n+1}^2$ in the physical subspace is given in the Appendix, Eq.~\eqref{eq:e_nn_simplified_supp}.

\section{Qubit vs Qudit Embedding}
\label{sec:qubit-vs-qudit-embedding}

To highlight the advantage of the native qudit encoding, we compare a single
symmetric Trotter step of our implementation with a qubit-based realization
obtained from standard exact decompositions. The resulting estimates are not
intended to be optimal lower bounds. Rather, they provide a conservative and
transparent benchmark for the overhead incurred when operations that are local
or two-body in the qudit Hilbert space are embedded into qubits.

We perform this gate-count comparison for the same finite system realized
experimentally and for the circuit structure of one symmetric Trotter step.
A finite-size feature of this instance is that the plaquettes at the
boundaries have a reduced local Hilbert space: because the boundary flux is
fixed, the corresponding qudits are effectively four-dimensional, while bulk
plaquettes are eight-dimensional. Thus, the nearest-neighbor interactions
appearing in the present string-breaking experiment act on one
four-dimensional and one eight-dimensional qudit, which map to two- and
three-qubit registers, respectively. Any additional plaquette added to the
bulk would correspond to a full eight-dimensional qudit. We nevertheless
restrict the gate-count comparison to the experimentally implemented system,
where the exact register dimensions and circuit structure are fixed.

For diagonal operations, we use the standard synthesis of arbitrary diagonal
$n$-qubit unitaries, requiring $2^n-2$ CNOT gates together with single-qubit
$R_z$ rotations~\cite{Bullock_2003}. For generic three-qubit gates, we use the
best known exact decomposition into 19 CNOT gates~\cite{Krol_2024}. A generic
two-qubit unitary is counted as 3 CNOT gates. These decompositions are
appropriate for the small registers considered here and provide a realistic
estimate of the entangling-gate overhead in a qubit implementation.

In the string-breaking simulations, each two-qudit interaction acts on one
four-dimensional and one eight-dimensional qudit, and therefore corresponds to
an effective five-qubit operation. Its diagonal component is implemented as an
arbitrary five-qubit diagonal unitary, requiring $2^5-2=30$ CNOT gates. The
four off-diagonal components are implemented by basis changes on the two- and
three-qubit registers, followed by diagonal entangling operations. Counting
3 CNOTs for each two-qubit basis change and 19 CNOTs for each three-qubit
basis change, one such two-qudit interaction requires
$30+4\times(3+19+30)$ CNOT gates. Since this interaction appears on two
neighbouring pairs in the experiment, the total two-qudit contribution is
\[
2\left[30 + 4\times(3+19+30)\right] = 476
\]
CNOT gates.

The remaining local qudit operations consist of plaquette terms and diagonal
energy terms. The local plaquette operations contribute four generic two-qubit
operations and two generic three-qubit operations, corresponding to
$4\times3+2\times19=50$ CNOT gates. The local diagonal energy terms contribute
a further 18 CNOT gates. Altogether, one symmetric Trotter step of the corresponding qubit circuit
requires
\[
476 + 50 + 18 = 544
\]
entangling gates. For the two symmetric Trotter steps used in the experiment,
this gives 1088 qubit entangling gates. By contrast, the native qudit
implementation uses 67 two-qudit entangling gates for the full two-step
evolution, corresponding to an overhead of approximately a factor of 16.

\section{Analytical solution in the first excited manifold for $J=1/2$ boundary charge}
\label{sec:analytical_solution}

In the perturbative regime $x \ll g^2$, the dynamics of the string fluctuations are restricted to the basis states of the first excited manifold (as shown in Fig.~\ref{fig:string_fluctuations}). We label these states as $\{ \ket{0}, \ldots, \ket{5} \}$, ordered from left to right within the $E_1$ manifold in Fig.~\ref{fig:string_fluctuations}. In this basis, the effective Hamiltonian from Eq.~\eqref{eq:fermionic_hopping} takes the form

\begin{equation}
H_\text{eff} = -\frac{x}{\sqrt{2}}
\begin{pmatrix}
0 & 1 & 0 & 0 & 0 & 0 \\
1 & 0 & 1 & 1 & 0 & 0 \\
0 & 1 & 0 & 0 & 1 & 0 \\
0 & 1 & 0 & 0 & 1 & 0 \\
0 & 0 & 1 & 1 & 0 & 1 \\
0 & 0 & 0 & 0 & 1 & 0
\end{pmatrix}
\!,
\end{equation}

Diagonalizing the Hamiltonian, we find analytical solutions for the evolution of the populations $P^{\pm}_{|n\rangle}(t) = \left|\bra{n}e^{-iH_\text{eff}t} \ket{\pm}\right|^2$  with initial states $\ket{\pm} = (\ket{2}\pm\ket{3})/\sqrt{2}$. For $\ket{+}$, they are given by
\begin{align}
    P^+_{\ket{1}}(t) = P^+_{\ket{4}}(t) &= \frac{2}{5} \sin^2(\Omega t), \\
    P^+_{\ket{2}}(t) =P^+_{ \ket{3}}(t) &= \frac{1}{50} \left(1 + 4 \cos(\Omega t) \right)^2, \\
    P^+_{\ket{0}}(t)=P^+_{\ket{5}}(t) &= \frac{2}{25}\left(\cos(\Omega t) - 1\right)^2
\end{align}
with the characteristic frequency
\begin{align}
    \Omega = \sqrt{\frac{5}{2}} x.
\end{align}
However, $\ket{-}$ is an eigenstate of the effective Hamiltonian and is therefore preserved in time. These expressions, valid to leading order in $x$, describe coherent oscillations within the first excited manifold and capture the interference-controlled population transfer between the dominant string configurations.

\section{Physical Subspace Reduction of the Electric Term}

The two-plaquette electric interaction $E_{n,n+1}^2$ acts on a local Hilbert space of dimension $8 \times 8 = 64$. However, due to the gauge constraints of the model, only 32 of these states belong to the physical subspace.  Defining projectors $P_{ij} = \ket{ij} \! \bra{ij}$, we write the original electric interaction from Eq. \eqref{eq:nonlocal_e_breaking} as
\begin{align}
\;&E_{n,n+1}^2 = \frac{g^2_{\perp}}{2} \times\\
&\frac{3}{4}\big(P_{01}+P_{10}+P_{03}+P_{12}+P_{21}+P_{23}+P_{35}+P_{37}\nonumber \\
&\quad +P_{40}+P_{42}+P_{54}+P_{56}+P_{65}+P_{67}+P_{74}+P_{76}\big) \nonumber \\[6pt]
&+2\big(P_{02}+P_{20}+P_{57}+P_{75}\big) \nonumber \\[6pt]
&+\big(P_{11}+P_{13}+P_{41}+P_{43}
+P_{34}+P_{36}+P_{64}+P_{66}\big) \nonumber \\[6pt]
&+\ket{11}\bra{34}+\ket{13}\bra{36}
+\ket{41}\bra{64}+\ket{43}\bra{66}
+\textit{h.c.}
\label{eq:e_nn_original_supp}
\end{align}

To simplify the term within the physical subspace, it is convenient to subtract a constant offset $3/4 \times \mathbf{I}$, which leaves the dynamics unchanged but yields a more compact representation. The operator then simplifies to
\begin{align}
\;&E_{n,n+1}^2 = \frac{g^2_{\perp}}{2} \times\\
&-\frac{3}{4}\big(P_{00}+P_{22}+P_{55}+P_{77}\big) \nonumber \\[6pt]
&+\frac{5}{4}\big(P_{02}+P_{20}+P_{57}+P_{75}\big) \nonumber \\[6pt]
&+\frac{1}{4}\big(P_{11}+P_{34}+P_{13}+P_{36}
+P_{41}+P_{64}+P_{43}+P_{66}\big) \nonumber \\[6pt]
&+\ket{11}\bra{34}+\ket{13}\bra{36}
+\ket{41}\bra{64}+\ket{43}\bra{66}
+\textit{h.c.}
\label{eq:e_nn_simplified_supp}
\end{align}

This form makes explicit that only a restricted set of transitions contributes to the dynamics. In particular, many of the terms that would arise from a generic two-qudit interaction are absent due to the gauge constraints. This sparsity directly translates into a reduced number of required entangling operations in the digital implementation, as unphysical transitions do not need to be generated or canceled at the circuit level.

\begin{figure*}[t]
    \centering
    \includegraphics[width=\textwidth]{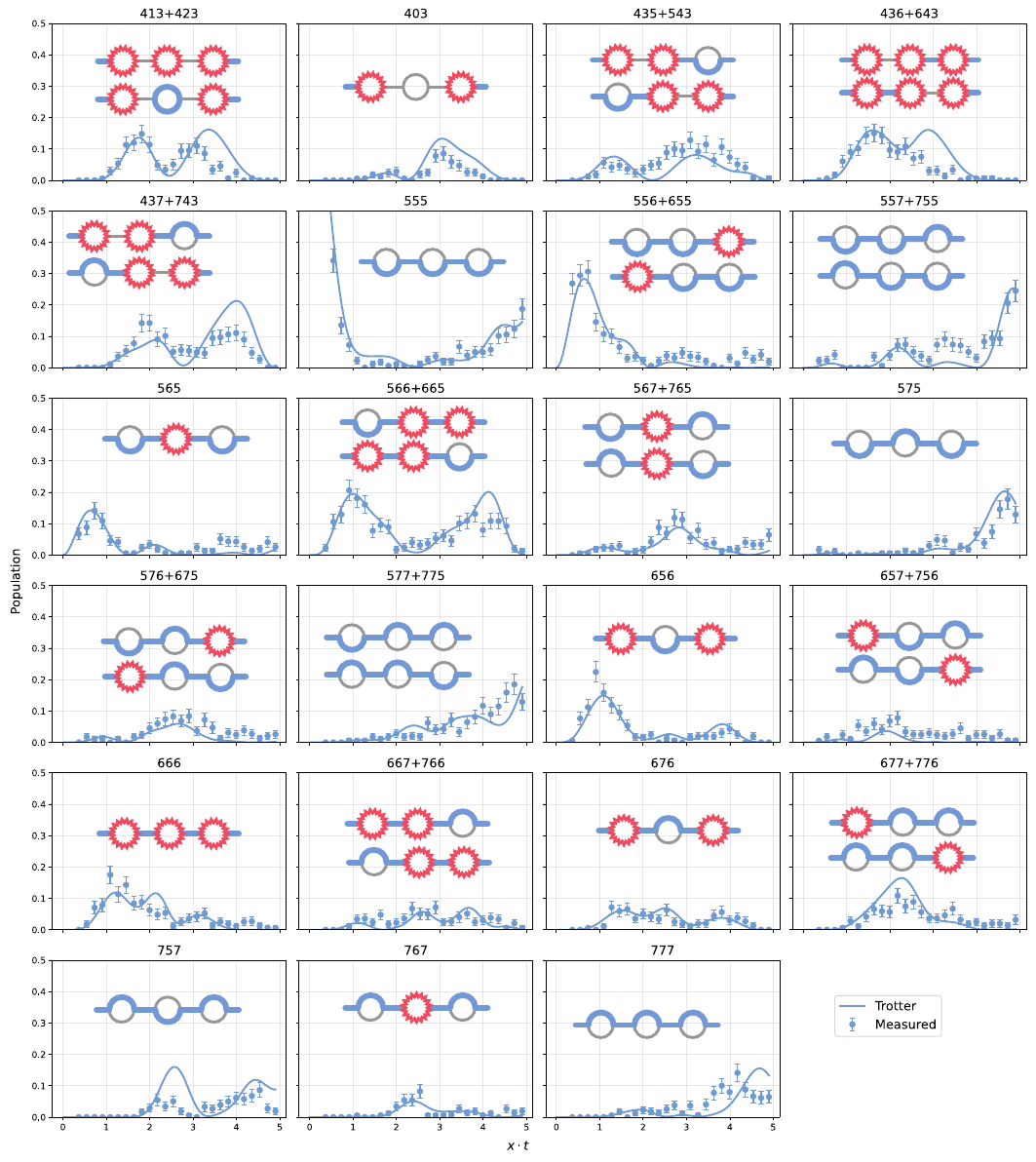}
    \caption{Example of time evolution showing all unique populations for $x=-0.78, g^2_h=2.8, g^2_v=2.0$. The populations of states $\ket{413}$ and $\ket{423}$ are summed as they are indistinguishable without running multiple measurements. Populations of states that show the same time-dynamics (due to symmetries) are summed up in simulation and experiment. The y-axis maximum is set to $0.5$ to make the comparison between simulation and experiment easier.}
    \label{fig:all_states_example}
\end{figure*}

\section{Full State Population Dynamics}

Figure~\ref{fig:all_states_example} shows the time evolution of all experimentally accessible populations for the parameters used in the main text. In contrast to the reduced observables presented in the main figures, this representation provides a complete view of the system dynamics within the measured subspace.

The data demonstrate that the observed dynamics are not confined to a small subset of states, but instead involve coherent population transfer across the full network of connected configurations. The agreement between experiment and simulation is maintained simultaneously for all populated states, providing strong evidence that the implemented dynamics capture the underlying many-body evolution beyond selected observables.

Due to readout constraints, the populations of the states $\ket{413}$ and $\ket{423}$ are combined, as they cannot be distinguished within a single measurement basis. Apart from this technical limitation, all populations are shown individually, except for symmetry-related states with identical time dynamics, whose populations are summed consistently in both simulation and experiment.

Overall, the global consistency across all populations highlights the coherent nature of the dynamics and supports the interpretation that the experimentally realized evolution faithfully reproduces the targeted Hamiltonian within the accessible timescales.

\section{Gate Count and Physicality}

\begin{figure}[t]
    \centering
    \includegraphics[width=\columnwidth]{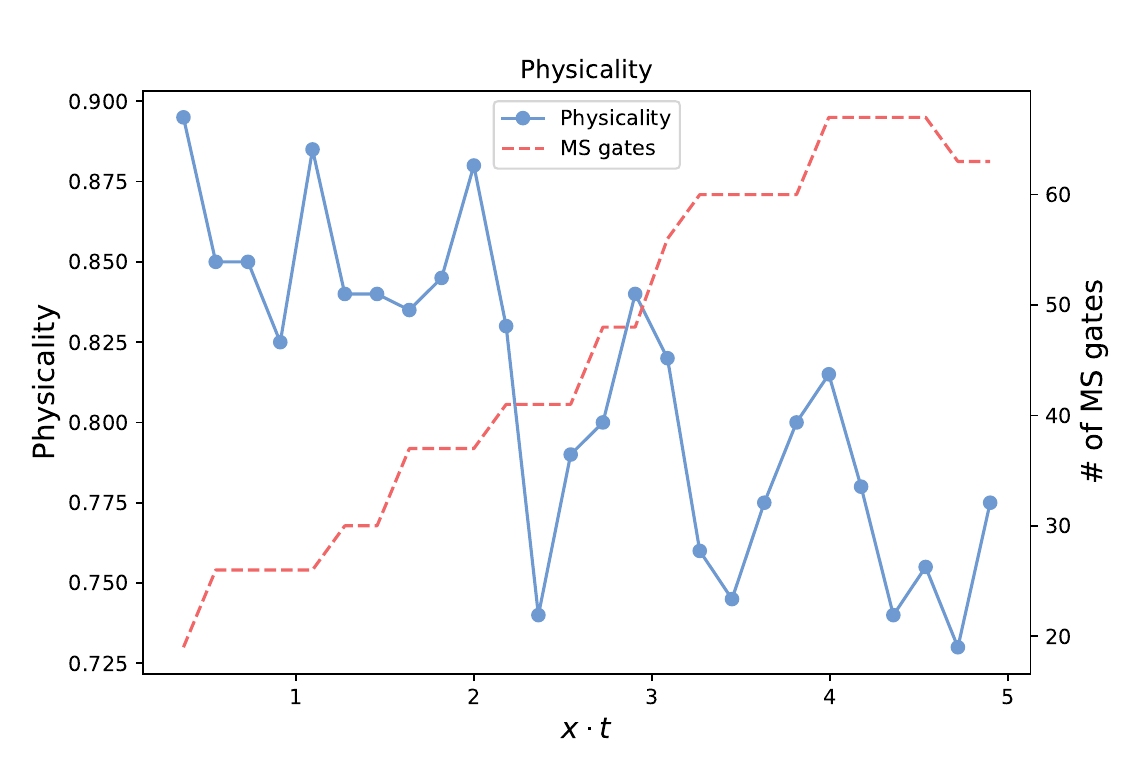}
    \caption{Example of physicality and gate count as a function of evolution time of the same simulations as in Fig.~\ref{fig:all_states_example}, for $x=-0.78, g^2_h=2.8, g^2_v=2.0$. The data shows an approximately linear trend even though we expect errors to accumulate exponentially with the number of MS gates. Since the MS gate is $4\pi$ periodic, the total number of MS gates can drop (as in the last two data points) if a particular rotation angle exceeds $2\pi$. In this case it becomes more efficient to rotate backwards.}
    \label{fig:physicality}
\end{figure}

Figure~\ref{fig:physicality} shows the total number of M{\o}lmer--S{\o}rensen (MS) entangling gates required for the simulations of Fig.~\ref{fig:all_states_example} as a function of the implemented evolution time. Despite the expectation that errors accumulate exponentially with circuit depth, the observed trend is approximately linear. 

Additionally, the intrinsic $4\pi$ periodicity of the MS interaction provides further optimization. For sufficiently large target angles, it becomes advantageous to implement rotations in the opposite direction, effectively reducing the total accumulated gate angle and, in some instances, the number of required entangling operations. This effect is visible in the non-monotonic behavior of the final data points in Fig.~\ref{fig:physicality}.

\end{document}